# A short account of a connection of Power Laws to the Information Entropy

Yaniv Dover

*Racah Institute of Physics, The Hebrew University, Jerusalem 91904, Israel*

*Tel: 972-02-6586786, e-mail: yanivd@phys.huji.ac.il*

**Abstract**

We use the formalism of 'Maximum Principle of Shannon's Entropy' to derive the general power law distribution function, using what seems to be a reasonable physical assumption, namely, the demand of a constant mean "internal order" (Boltzmann Entropy) of a complex, self interacting, self organized system.

Since the Shannon entropy is equivalent to the Boltzmann's entropy under equilibrium, non interacting conditions, we interpret this result as the complex system making use of its intra-interactions and its non equilibrium in order to keep the *equilibrium* Boltzmann's entropy constant on the average, thus enabling it an advantage at surviving over less ordered systems, i.e. hinting towards an "Evolution of Structure".

We then demonstrate the formalism using a toy model to explain the power laws observed in Cities' populations and show how Zipf's law comes out as a natural special point of the model. We also suggest further directions of theory.

PACS numbers **89.70.+c, 89.75.-k, 02.50.-r, 05.65.+b**.

Keywords: Power Laws, Information theory, Statistical physics, Dynamical systems, Self Organizing systems.

## 1. Introduction

Since the general acceptance that Power Law Probability Distribution Functions are abundant in nature and so are interesting enough, several theoretical attempts have been made to derive a primal principle which will produce these power law distribution functions using known and "natural" axioms.

Some of these were the "Self Organized Criticality" theory [1], "The Multiplicative Formalism" (e.g. Solomon [6]), or Tsallis [2], who assumed a new specific kind of non extensive Entropy from which to derive in a "Maximum Entropy Principle", these Power Law distributions. These are only a part of the list.

We suggest here the use of a primal physical/mathematical principle, "Shannon's Entropy", which is widely used and accepted. This principle will be used to derive these power law distribution functions, combined with some physical assumptions. The motivation for the use of this formalism is the thought that existing functioning systems in nature have some evolutionary, naturally selected mechanisms to improve self organization within them which bears the insignia of power laws.

If indeed this mathematical/physical frame is agreed upon by experiment, then this formalism could supply the quantitative phenomenological relation of the macroscopic power law distribution functions to another macroscopic phenomena, "internal order" which could be modeled in a simpler, more intuitive manner and hint to the origin of the accumulated data pointing towards Power Laws' abundance in nature.

In other words, we could assign the Shannon's statistical mechanics' techniques to those systems which are mostly highly self interacting and out of equilibrium and are difficult to model otherwise, exploiting the remarkable connection (see below) of power law distribution functions to "internal order/entropy" as, at least, is depicted by this very simple Shannon formalism.

What systems are we talking about?
It is known for a long time that in order to exist, some "biological" systems maintain a certain kind of "internal order", struggling constantly against the second law of Thermodynamics for survival. A general, crude, example could be the "living organism" which struggles to maintain energy and temperature against the ongoing heat dissipation to the environment's "heat bath".

One could also suspect that biological information networks that need a certain minimal amount of information stored within them in order to justify their existence and resource consumption, to work towards conserving their inner structure (which could be assigned quantifiable order) and thus join the category above.

Of course, further investigation should be made in order to identify these systems as valid for this discussion.

Some less obvious examples can be taken from "sociological structures" which could be analogical to physical systems. We will not deal here with the validity of the analogy but use one simple example for pedagogical reasons, namely the 'Country', 'City', 'Citizen' set. Where we mean, 'System', 'Subsystem' and 'Agent' respectively. This last example will be stressed below, where it will be shown under our formalism that the probability of having a city with $N$ inhabitants is proportional to $N^{-1-a}$, i.e. a power law dependence, with special meaning for the exponent, just as is evident in real life data[8].

We will note in passing that a similar basic mathematical formalism was already used in an economics related article [3] which after articulating the
formalism herein we found following an extensive literature search. Although the same basic mathematical manipulation was used, it wasn't taken to the general direction we are proposing here.

## 2. Fixed "internal entropy/order" (The Model)

Say a system is composed of subsystems, interacting with each other in a complicated manner. These subsystems are composed each of agents (e.g a City, in a Country, is composed of 'individuals'- the microscopic building blocks) which interact with each other in a complicated manner also.

Say that the survival of the system is dependent upon the "internal order" of the system, where "internal order" will be defined as the mean of the Boltzmann's Entropy over all the subsystems (see Eq (4)), i.e. we suggest that the survival of the system is dependent somehow on keeping the *mean Boltzmann's entropy* of the system, fixed, either in time or fixed to a certain desired value which could change over time.

The importance of order to an organic/social system is apparent, when you consider for example the crudest example of a cell, which must gather resources crucial for its survival, fighting against their tendency to spread spatially as the second law of thermodynamics implies.

## 3. Formulation

Deriving from the last section, we propose a restriction in the form:

$$<s> \overset{!}{=} const \qquad (3a)$$

Where we denote $<s>$, as the *mean Boltzmann entropy* of the system over all subsystems (which in the theory means the average over an ensemble of subsystems). The passage to the required constancy of entropy in time could be understood from something similar to an ergodic assumption, where the mean over ensembles should be close to the mean over a long period of time. This of course is justifiable for large ensembles or long time scales, for example and it is assumed here that smaller ensembles require some sort of corrections but do not change the essence of the theory.

Restriction (3a) could also be understood as the mean Entropy with a time dependence but at each instant of time it is not the natural entropy of the system in an independent manner but some value which is determined by external conditions.

So we define the Boltzmann Entropy:

$$s_i \equiv -\ln(\Omega_i(x)) \quad (3b),$$

Working with units where $k_B$ is unity.

$\Omega_i(x)$ being the number of configuration states (Statistical Weight) for the subsystem $i$ with parameter $x$ and so (3a) becomes:

$$<s> = \sum_i p_i(\Omega(x)) \cdot s_i(\Omega(x)) = -\sum_i p_i(\Omega(x)) \cdot \ln(\Omega(x)) \overset{!}{=} const \quad (4)$$

Where the mean is over all 'subsystems'/ensemble (here the minus is inserted into the constant's definition so that (5) could be written as it is) and only one independent parameter ($x$) was taken for the sake simplicity.

Now, with this restriction (4) over the distribution we seek and the normalization condition restriction (8), we write the following Shannon Lagrangian:

$$L \equiv -\sum_i p_i(\Omega_i(x)) \cdot \ln(p_i(\Omega_i(x))) - l_1 \cdot \left[\sum_i p_i(\Omega_i(x)) \cdot \ln(\Omega_i(x)) - <s>\right] - l_2 \cdot \left[\sum_i p_i(\Omega_i(x)) - 1\right] \quad (5)$$

Where we denote the constant entropy average as simply $<s>$.

Extremizing (maximizing actually, since the positive PDF makes the second variational deriv. be negative) this expression, in order to arrive at the distribution functions which correspond to restriction (4) and (8):

$$\frac{\partial L}{\partial p_i(\Omega_i(x))} \equiv -1 - \ln(p_i(\Omega_i(x))) - l_1 \ln(\Omega_i(x)) - l_2 = 0 \quad (6)$$

Which gives us the distribution functions:

$$p_i(\Omega_i(x)) = p(x) = e^{-1-l_2} \cdot \Omega_i^{-l_1} \quad (7)$$

Now, the normalizing condition:

$$\sum_i p_i(\Omega(x)) = 1 \quad (8)$$

Gives us an expression for the $l_2$ dependent factor in (7):

$$\sum_i e^{-1-l_2} \cdot \Omega_i^{-l_1} = 1 \quad (9)$$

i.e.:

$$e^{-1-l_2} = \frac{1}{\sum_i \Omega_i^{-l_1}} \equiv \frac{1}{Z} \quad (10)$$

This defines the Partition function, $Z$ in this case. So we have (7):

$$p_i(\Omega_i(x)) = \frac{1}{Z} \cdot \Omega_i^{-l_1} \quad (11)$$

This is a power law of the probabilities in the statistical weight of the sub systems.
In other words, a system that has the restriction of normalized probabilities, produces the Gibbs distribution, this is known for a long time[7], while in our case, since we've added another restriction, namely the restriction of constant Boltzmann's entropy (4), the distribution has become a power law in the statistical weight of the system (11).

A derivation for a continuous probability distribution is trivial and does not change theory in essence and will not be given here.

Before we continue, we'll just mark here some troublesome points which will be addressed in the "Discussion" section below.

One should notice that only a "power dependence" of the number of configuration states on any parameter will produce a power law distribution functions, this is indeed a limitation over the possible physical models.

Another limitation is that, $<s>$ or interchangeably $l$ (Since they are connected by restriction (4)), are "external parameters" and not specified in this formalism. Although there is an obvious fundamental connection of $<s>$ to that "internal order" of the system, it seems that one should derive it in a more specific model from further considerations natural to that model.

## 4. Example of Application, a Toy Model for Cities' Distribution:

Say we have a 'Super Group' (a System) of agents; we shall call "*Country*", where the 'Super' means it is composed of sub groups (Sub systems) which we will call "*Cities*". The question we may want to pose is, how those agents will divide into Cities in a manner that will enable those Cities to maintain their "internal order" constant but still fall under the laws of nature, i.e. maximize Shannon's entropy.
( one may also note that this average 'internal order' of 'Cities' is actually simply proportional to the total 'internal order' of the "*Country*").

In order to quantify the "internal order" we mentioned, we model a *City* as a 'network' of interconnected "individuals/agents". We can use the links among agents/nodes to define the number of states it could take (with given number of agents, $N$), which we suppose is connected to its ability to function and therefore "survive".
Say that the total number of crucial intra connections needed in order to maintain its inflow of essential resources (e.g. information, supplies etc.), is another constant of a system.
This number, we assume, is different for each Country, but constant within it, reflecting the external conditions a Country/Central Government provides from which Cities are created and draw their existence over time.
Lets denote this number as '$c$'. It will be called the effective number of 'vital connections' or just 'vital connections'.
If we assume that the "internal order" or logarithm of "number of possible configuration states" are in this meaning related to the layout of these connections we will count in a crude simplistic manner the number of possible states per City of $N$ agents to be the total possible number of configurations of connections among agents, provided it will amount to a total of '$c$' connections.
A point should be made here, that this '$c$' is not in fact the number of connections responsible for the possible size of the City and representing its infra structure which should be different for different sized Cities, but actually is the number of connections vital for a City's *existence*, which should therefore be effectively constant in a specific Country for large and small Cities as one, since it reflects the external conditions a Country/Central Government provides for the creation of a City as a functioning, surviving, structure.
The total number of intra, double sided connections available in a given City of $N$ inhabitants is: $\frac{N^2-N}{2}$.

So the count of configurations of $c$ connections out of $\frac{N^2-N}{2}$ links is:

$$\Omega(N) = \binom{\frac{N^2-N}{2}}{c} = \frac{(\frac{N^2-N}{2})!}{(\frac{N^2-N}{2}-c)! \, c!} = \frac{(\frac{N^2-N}{2}) \cdot (\frac{N^2-N}{2}-1)...(\frac{N^2-N}{2}-c+1))}{c!} \approx \frac{N^{2 \cdot c}}{2^c \cdot c!} \quad (12)$$

Where we took $N \gg c$ for analytical simplicity (which is reasonable in real cases of large networks with small compared 'number of vital connections'). Also we neglected the linear term, which seems quite reasonable for the number of agents we're dealing with.

One must note that this approximation becomes less applicable for very small $N$, so we assume as in most of the highly populated systems that the minimum value for a functioning City could be large enough, as it is in the experimental results we quote [8].

We also assumed, again for simplicity, that a certain agent has equal probabilities of obtaining a link with any other agent, which is an assumption that should be considered over the relevant time scales, i.e. time scales of Cities' creation, destruction and survival.

So the total number of possible states in City $i$ should be (rewriting (12)):

$$\Omega_i(N) = \frac{N_i^{2 \cdot c}}{2^c \cdot c!}. \quad (13)$$

Putting this relation (13) in equation (11):

$$p_i(x) = \frac{1}{Z} \cdot \left( \frac{1}{2^c \cdot c!} \right)^{-\mathbf{l}} \cdot N_i^{-2 \cdot c \cdot \mathbf{l}} \quad (14a)$$

Or

$$p_i(x) = \frac{N_i^{-2 \cdot c \cdot \mathbf{l}}}{\sum_i N_i^{-2 \cdot c \cdot \mathbf{l}}} \quad (14b)$$

Which is a power law, such as the power laws seen in Cities' populations [5] and many other systems [1]-[3],[5]-[6], where the exponent usually denoted as $\mathbf{a}$ is related to $\mathbf{l}, c$ by:

$$-1 - \mathbf{a} \equiv -2 \cdot \mathbf{l} \cdot c \equiv -\frac{2 \cdot c}{\mathbf{h}} \quad (15a).$$

Where :

$$\mathbf{h} \equiv \frac{1}{\mathbf{l}} \quad (15b)$$

In order to make some observations regarding the relation of our simple toy model to the existing data, we look at $c$ using (15a) and (15b):

$$c = \mathbf{h} \cdot \left( \frac{\mathbf{a}+1}{2} \right) \quad (16)$$

i.e. here we deduct the meaning of $\mathbf{l}$ ($\mathbf{h}$), using its algebraic role, i.e. to inhibit the role of '$c$' in the exponent of (14b). It is simpler to see this if we write the exponent in (14b) instead of $-2 \cdot \mathbf{l} \cdot c$ as $-\frac{c}{\left( \mathbf{h}/2 \right)}$.

Since $\mathbf{h}$ is a trait of the theory and hasn't come from considerations of the specific details of the system, we assume that it is a global quantity (i.e. independent of a specific Country).

We interpret $h$ as the number of vital connections dictated by the nature of the theory itself or something else external to the systems in question. This number could in fact be the *global* number of intra-connections necessary to create a City or keep it in existence. Another possible way of understanding $h$ through Zipf's law is given below. Of course further understanding of it is required through a more detailed theory.

If it is such a global number of vital connections, then the exponent ($a$) expresses the number of vital connections a certain Country has in relation to that global number, as could be seen from (16).
We can see that in this theory the Zipf's rule comes out naturally as a special point, without *a priori* assumptions. This is since when $a \to 1$, $c \to h$, i.e. the number of vital connections in a certain Country equals that of the global number when Zipf's law is fulfilled. In other words, Zipf's law as understood in our model could be stated as the physical desire of Cities' structures to attain the needed number of 'vital connections' in order to survive.
We are therefore obliged to say that this relation clearly has a deeper meaning in that case in relation to the Country's economical, sociological or political state, since the data's readings below, show trends.
Using the Data in [8], it is apparent that the average of exponents ($a$) in Europe, Oceania and North America is over 1 while the average for Asia, Africa and South America is below 1.
In terms of our model, this could mean that in those highly industrialized Countries, the number of vital connections needed to create a City is higher then in those less industrialized. Perhaps due to higher complexity of the urban network that is required of a City in these industrialized parts. Deeper understanding in the economical, sociological or political context of this finding should be deducted from a more detailed model and detailed research motivated by the simple model entailed here above. Such an understanding could teach us a great deal about human social structures dynamics. Although this model is very simple and demands further discussion regarding the validity of the approximations taken, still intuitively, these results look promising in the further understanding of power laws and the systems, which exhibit them.

## 5. Discussion

Although there is a debate in the literature over the validity of such a theory, emerging from "Information Theory" considerations, where the resulting PDF is too sensitive to the "arbitrary" restrictions imposed on the system, in our formalism there is a sound physical basis for said restrictions, namely the conservation of "internal order"/Boltzmann Entropy. This basis, we think, narrows the arbitrariness in a profound way.

The Approximation made here in order to achieve a power law distribution functions, need further discussion and handling of definitions, since we are dealing with such complicated systems. But we suspect that the essential features of this theory will not be changed drastically under reconsiderations.

One could also point out after observing that the functional form of the key quantity, namely the "statistical weight of a subsystem", $\Omega_i(x)$ is constrained, that perhaps this formalism may not always produce a power law, e.g. if the functional dependent of $\Omega_i(x)$ in it's parameter, $x$, is exponential, then the resulting PDF will be exponential also. To this, one may answer that not ALL "self organized" "internal ordered" systems in "Nature" exhibit power law distribution functions and so the formalism isn't obviously dismissed but further gains an opportunity to validate it's compatibility through defining testable arguments for when one should expect a power law distribution functions and when one should not.

Regarding the external parameter $<s>$, we can only suggest the formulation of some microscopic or phenomenological theory that will produce a connection of this parameter to the specific system's details. One could also assume that the theory suggested here is loyal to experiment and use it to explore the nature of the obvious connection of $<s>$ to the "internal order" of the system.
Another option is to use $l(h)$ and restriction (4) in order to derive the meaning of $<s>$, assuming the meaning of $l(h)$ is known (as we believe is in the Toy model's case).

# 6. Conclusions

We have shown here, using some of the very basics of the well known "Information Theory", a starting point from which to begin partially phenomenological investigations of Complex, self interacting systems exhibiting Power Law distribution functions.

This could be explained in an intuitive way. In fact, there is nothing new in the statement that the Shannon's entropy and Boltzmann's entropy are two different quantities, where the last is a special case of the first [7].

Actually, what we have shown here, is that if one virtually separates Shannon's entropy into two categories of entropy, the so called "local" entropy of each subsystem (e.g. *Cities*) i.e. the Boltzmann entropy of each subsystem and the entropy which is "everything else" which we can only deduce, is a measure of the entropy of interactions of these subsystems amongst themselves, the interactions amongst their agents (inter-Cities interactions), the deviation from equilibrium and the so called "information" stored in the system, one can observe non standards behaviors such as power law distribution functions.

In other words, we could say that the system has a "local" entropy and a "non local" entropy divided as explained above. Since the system somehow works towards maintaining the useful "local" entropy at bay (what we called 'internal order'), it 'uses' the "interaction" or "non local" entropy in a manner that will allow it to do so, i.e. fix the Boltzmann Entropy.

Since systems that manage to do so, have probably better chance of survival, it is possible that this is one of the preferred types of structures in our times. In other words, we demonstrated some sort of "Evolution of social/organic Structure".

What we've shown here is that in the case that the system abides by the limitations we mentioned above, it bears the insignia of the power law distribution functions.

We have also given a very simple example of a model that belongs to a group of models, which display that Power Law PDF. This model gives an explanation for Zipf's law and also for the deviations from Zipf's law.

This model could also be useful in other organic/social structured phenomenon in nature for which deeper physical understanding is required.

I wish to thank Prof. Sorin Solomon, under whose inspiration this research was possible and both him and Uri Hershberg for useful discussions.